\documentclass[amsmath,amssymb,superscriptaddress,showpacs]{revtex4-2}

\usepackage{graphicx}
\usepackage[caption=false]{subfig}
\usepackage{dcolumn}
\usepackage{bm}
\usepackage[colorlinks=true,citecolor=blue,urlcolor=blue,linkcolor=red,bookmarks=true]{hyperref}
\usepackage{placeins}
\usepackage{multirow}

\begin{document}

\title{Topological signatures in Kerr-Sen AdS black hole thermodynamics }
\author{Mohd Rehan} 
\email{rk1958227@gmail.com} 
\affiliation{Centre for Theoretical Physics, Jamia Millia Islamia, New Delhi 110025, India}
\author{Md Sabir Ali} 
\email{alimd.sabir3@gmail.com, sabirali@mahishadalrajcollege.ac.in} 
\affiliation{Department of Physics, Mahishadal Raj College, Purba Medinipur, West Bengal 721628, India}
\affiliation{Institute of Theoretical Physics \&
Research Center of Gravitation, Lanzhou University, Lanzhou 730000, China}
\affiliation{School of Physical Science and Technology, Lanzhou University, Lanzhou 730000, China}
\author{Sushant G. Ghosh} \email{sghosh2@jmi.ac.in}
\affiliation{Centre for Theoretical Physics, Jamia Millia Islamia, New Delhi 110025, India}
\affiliation{Astrophysics and Cosmology Research Unit, School of Mathematics, Statistics and Computer Science, University of KwaZulu-Natal, Private Bag X54001, Durban 4000, South Africa}
\date{\today}

\begin{abstract}
Black hole thermodynamics and topology have emerged as a strong foundation for a coordinate-independent understanding of phase transitions. Using both Duan's topological current theory and a novel complex residue method, we perform a topological study of the Kerr-Sen AdS black hole arising in heterotic string theory. In turn, we find the zero points corresponding to on-shell black hole states and calculate their winding numbers to find the global topological charge by building the generalized off-shell free energy and examining the corresponding vector field in a parametric space. Our analysis reveals that the Kerr-Sen AdS black hole exhibits three distinct thermodynamic phases---small, intermediate, and large black hole branches---characterized by critical points with winding numbers $+1$, $-1$, and $+1$ respectively, culminating in a total topological charge $W = +1$. Significantly, this topological number remains invariant under variations of the dilaton charge parameter, indicating that the dilaton field does not alter the fundamental topological class established for Kerr-AdS and RN-AdS black holes. However, the rotation parameter proves crucial in determining the phase structure and the emergence of multiple critical points. We systematically examine three limiting configurations: the full Kerr-Sen AdS spacetime, the GMGHS AdS limit ($a = 0$), and the asymptotically flat Kerr-Sen case ($\Lambda = 0$). The GMGHS AdS black hole yields a vanishing total topological charge with winding numbers $+1$ and $-1$ cancelling each other, while the Kerr-Sen black hole ($\Lambda = 0$) similarly exhibits $W = 0$ despite possessing two distinct phase branches.  Furthermore, we propose a novel approach by analytically continuing the thermodynamic characterisation to the complex plane. The characterized complex function, derived from the off-shell Gibbs free energy, possesses isolated singular points whose residues directly encode the winding numbers.  Our results indicate that topology offers deep insights into black hole phase transitions, with potential implications for understanding holographic dualities.
\end{abstract}

\maketitle

\vspace{0.5cm}
\section{Introduction}
Black holes have developed from an exact solution of Einstein's general relativity (GR) to observationally accessible astrophysical objects, with the LIGO/Virgo collaborations detecting gravitational waves from black hole mergers \cite{LIGOScientific:2016vlm, LIGOScientific:2016sjg} and the Event Horizon Telescope (EHT) providing the first direct images of black hole shadows of M87* and Sgr A* \cite{EventHorizonTelescope:2019ggy, EventHorizonTelescope:2022wkp}. The EHT and LIGO/Virgo observations have opened a window for an era of precision strong gravity, enabling exceptional tests of gravitational theories in the extreme curvature regime.
The current decade is celebrated as the dawn of precision strong gravity. The observational data from  LIGO/Virgo/LISA and EHT/ngEHT are being tested, particularly, the true nature of the astrophysical black holes. To see the unseen and to hear the echoes from the darkest side of the universe are within our reach. Hence, the gravitational wave detections of black hole mergers through the ringdown phase and shadow imaging of the black holes via electromagnetic observations are expected to provide us with hitherto inaccessible information on the strong field regime of gravity. Theoretically, it is anticipated that exploring the topology of black holes to uncover the true nature of gravity can be done  \textit{vis-\`a-vis} knowing the topological structure of light rings, timelike circular orbits \cite{Cunha:2017qtt,Cunha:2020azh}, and thermodynamic properties \cite{Wei:2022dzw, Wei:2021vdx,Rehan:2024dsg}.   

The thermodynamic interpretation of black holes stands as one of the most profound conceptual developments in theoretical physics. The  four laws of black hole mechanics \cite{Bardeen:1973gs} and the Hawking radiation \cite{Hawking:1974sw, Hawking:1975vcx} indicated that black holes possess temperature and entropy, fundamentally connecting gravitation, quantum mechanics, and thermodynamics. The thermodynamic analogy naturally appears in anti-de Sitter (AdS) spacetimes, where black holes exhibit rich phase structures reminiscent of van der Waals liquid-gas systems \cite{Hawking:1982dh}. While the Hawking-Page phase transition between thermal AdS space and large Schwarzschild-AdS black holes \cite{Hawking:1982dh} marked a remarkable moment, establishing black hole thermodynamics as a fruitful ground for exploring gauge/gravity dualities. The phase transition behaviour of AdS black holes and their topological classification using the generalized off-shell free energy approach provide valuable insights into the fundamental nature of gravity\cite{Li:2021tpu,Li:2022ylz,Liu:2023sbf,Li:2021zep,Li:2023men}.

In recent years, the investigation of black hole thermodynamics has been revolutionised by the application of topological methods and inspiration from Duan's $\phi$-mapping topological current theory \cite{Duan:1979ucg, 2001MPLA...16.2483D, Fu:2000pb}, Wei, Liu, and Mann \cite{Wei:2022dzw, Wei:2022mzv} introduced a novel framework wherein black hole solutions are treated as topological defects in a thermodynamic parameter space. To do this, they construct a generalised off-shell free energy and study a vector field associated with it. The zero points of this vector field are exactly the on-shell black hole states and have a winding number, a topological charge. Adding these winding numbers gives a topological number that describes the whole system. Remarkably, this number doesn't change when thermodynamic parameters vary smoothly—it's a universal property. The same approach has also been used to study timelike circular orbits around stationary black holes \cite{Wei:2022mzv}.

The topological framework has been systematically applied across a diverse range of black hole solutions. Studies have explored vacuum AdS solutions in classical GR  \cite{Yerra:2022coh}, regular black holes including Bardeen and Hayward-AdS solutions \cite{Sadeghi:2023aii, Fang:2022rsb}, Einstein-Gauss-Bonnet AdS black holes \cite{Liu:2023rqj}, Lovelock gravity \cite{Bai:2022klw}, and Born-Infeld AdS black holes \cite{Ali:2023jox}. The rotating spacetimes \cite{Wu:2022whe, Wu:2023sue} have revealed that the spin significantly influences topological classifications, with Kerr-Sen-AdS black holes exhibiting superentropic behaviour \cite{Wu:2020mby}. The thermodynamics of Kerr-Sen-AdS black holes has been further investigated in the restricted phase space formalism \cite{Ali:2023ppg, Gao:2021xtt, Wu:2020cgf}.

Similarly, Fang \cite{Fang:2022rsb} proposed a complementary approach that applies complex residue techniques: by analytically continuing thermodynamic functions into the complex plane, one can construct characterized functions whose isolated singularities correspond to thermodynamic defects. It turns out that the residues at these poles encode winding numbers, providing a mathematically elegant method that reproduces and extends Duan's topological approach. Recent work has extended this approach to the restricted phase space thermodynamics of Kerr-Sen-AdS black holes \cite{Hazarika:2024dex}. 
Establishing the correspondence between Duan's topological current method and residue techniques helps us distinguish the creation and annihilation profiles of black hole phase transitions near critical points. Such methods have been employed for a class of black hole solutions where the creation/annihilation and the critical points are categorized by coefficient analysis of the Laurent series of the complexified off-shell Gibbs free energy.\\  
Of particular interest, in a more recent version of the geometric interpretation \cite{Wei:2019yvs,Wei:2015iwa,Wei:2019uqg,NaveenaKumara:2020biu}, black hole phase structures are probed through the investigation of topological classifications based on the idea of Duan's $\phi$-mapping topological current and the corresponding winding number defined in an auxiliary parameter space\cite{Wei:2021vdx,Wei:2022dzw}. When treating the thermal phase behavior of black hole solutions as various topological defects, we classify them into three distinct categories, each corresponding to a different topological charge and associated topological number. Hence, the classification of black hole solutions as topological defects and the investigation of phase transition behavior at critical points through a temperature-dependent function categorize topological charges broadly into two distinct subclasses: one corresponding to conventional critical points, while the other is associated with novel phase transitions. \\ 
Despite significant investigations, the topology of the AdS black holes still remains elusive. The rationale behind such a study incorporates topological numbers that bear universal features independent of the parameters characterizing the black hole. Hence, the topology plays a pivotal role in deciphering the nature of black holes and gravitational theory. Due to its versatility and naturalness, the topological framework has gained rapid acceptance owing to its conceptual simplicity and adaptability to diverse gravitational theories. Consequently, ever since its inception, the topological numbers and their classifications have been explored in a wide range of known black hole solutions, e.g., the vacuum/electro-vacuum AdS solutions in classical general relativity \cite{Yerra:2022coh}, the regular AdS black hole solutions \cite{Sadeghi:2023aii, Fang:2022rsb,Sadeghi:2024krq}, the Einstein-Gauss-Bonnet AdS black holes \cite{Liu:2022aqt}, the black holes in Lovelock gravity \cite{Bai:2022klw}, Born-Infeld AdS black holes \cite{Ali:2023jox}. Later, such studies have been extended to the rotating spacetimes \cite{Wu:2022whe, Wu:2023sue, Wu:2020cgf}. Since the topological classifications of the rotating AdS black holes are still in their early phase, they deserve to be deeply analyzed. Further, the rotating AdS black holes are fundamentally connected to the gauge/gravity dualities, as they shed light on the nature of quantum gravity. Such notions motivate us to study the thermodynamic topology of the Kerr-Sen AdS black holes in their own right, as it provides a global, coordinate-independent framework for our basic understanding of black hole thermodynamics during phase transitions--especially in light of its emergence from string theory. \\
Motivated by this, we handle the above questions via a topological analysis of Kerr-Sen AdS black holes. We examine three distinct configurations: the full Kerr-Sen AdS spacetime, its non-rotating limit (GMGHS AdS black hole), and the asymptotically flat Kerr-Sen case ($\Lambda = 0$). We use Duan's $\phi$-mapping topological current theory to compute winding numbers and global topological charges for each configuration, elucidating the roles of rotation, dilaton charge, and cosmological constant. We further compute the angular deflection of the unit vector field to characterise the topological charge of each critical point. Additionally, we apply the complex residue technique to this system, demonstrating its ability to reproduce and extend the topological classification while offering new insights into the nature of critical points.

The organization of the paper is as follows. In Sec.\ref{II}, we briefly overview the Kerr-Sen AdS black holes and their thermodynamic quantities. Using the definition of the off-shell Gibbs free energy, we identify the topological charges and their possible classifications based on Duan's topological mapping theory. In Sec.~\ref{III}, we analyze our results in the context of complex residue techniques by defining a characteristic function constructed from the off-shell Gibbs free energy of a Kerr-Sen black hole. In Sec.~\ref{IV}, we discuss our results and conclude the paper.\\
\section{Kerr-Sen black hole and its Topology}
\label{II}
The Kerr-Sen AdS black hole is a rotating black hole spacetime 
arise from the low-energy effective action of heterotic string theory, incorporating both a dilaton field and a negative cosmological constant. The Kerr-Sen AdS black hole is a generalization of the original Kerr-Sen spacetime \cite{Sen:1992ua} to asymptotically anti-de Sitter backgrounds. The Kerr-Sen AdS black hole is particularly significant in the context of black hole thermodynamics, as the presence of the cosmological constant introduces a rich phase structure similar to van der Waals liquid-gas systems. 
At the same time, its string theory origin provides a natural framework for exploring corrections to Einstein gravity and their implications for the gauge/gravity duality. In the usual Boyer-Lindquist coordinates, the Kerr-Sen AdS metric takes the form \cite{Wu:2020cgf, Ali:2023ppg, Wu:2021mkk}
\begin{equation}
\label{Kerr}
\begin{aligned}
ds^{2}
&=
-\frac{\Delta_{r}}{\rho^{2}}
\left(
dt-\frac{a\sin^{2}\theta}{\Xi}\, d\phi
\right)^{2}
+\frac{\rho^{2}}{\Delta_{r}}\,dr^{2}
+\frac{\rho^{2}}{\Delta_{\theta}}\,d\theta^{2}
\\
&\quad
+\frac{\sin^{2}\theta\,\Delta_{\theta}}{\rho^{2}}
\left(
a\,dt-\frac{r^{2}+2br+a^{2}}{\Xi}\,d\phi
\right)^{2}.
\end{aligned}
\end{equation}

where
\begin{align}
\rho^{2}
&=
r^{2}+2br+a^{2}\cos^{2}\theta,
\\[4pt]
\Delta_{r}
&=
\left(r^{2}+2br+a^{2}\right)
\left(1+\frac{r^{2}+2br}{\ell^{2}}\right)
-2Gmr,
\\[4pt]
\Xi
&=
1-\frac{a^{2}}{\ell^{2}},
\qquad
\Delta_{\theta}
=
1-\frac{a^{2}}{\ell^{2}}\cos^{2}\theta .
\end{align}

\begin{eqnarray}
\label{Therm}
&&M = \frac{m}{\Xi^2} \, , \quad J = \frac{ma}{\Xi^2} \, , \quad
Q= \frac{q}{\Xi}
\end{eqnarray}
where the mass parameter $m$ can be obtained from the condition $\Delta_r(r=r_h)=0$ with $r_h$ denoting the radius of the event horizon. Consequently, the mass $M$ and angular momentum $J$ can be expressed in terms of $a,\;r_h,\;G$. The thermodynamic quantities associated with the above solution (\ref{Kerr}) can be derived
using the standard methods, and are given by the following explicit expressions \cite{Ali:2023ppg,Wu:2020cgf} 
\begin{eqnarray}
\label{mJ}
&& M=\frac{1}{2\Xi^2 G r_h}\left(r_h^{2}+2br_h+a^{2}\right)\left(1+\frac{r_h^{2}+2br_h}{\ell^{2}}\right),\; \\ && J=\frac{a\left(r_h^{2}+2br_h+a^{2}\right)}{2\Xi^2 G r_h}\left(1+\frac{r_h^{2}+2br_h}{\ell^{2}}\right)
\end{eqnarray}
The corresponding expressions for Kerr-AdS black holes are obtained when $b=0$ \cite{Gao:2021xtt},
\begin{eqnarray}
\label{mK}
&& M=\frac{1}{2\Xi^2 G r_h}\left(r_h^2+a^2+\frac{r_h^{4}}{\ell^{2}}+\frac{a^2r_h^{2}}{\ell^{2}}\right),\; \\ && J=\frac{a}{2\Xi^2 G r_h}\left(r_h^2+a^2+\frac{r_h^{4}}{\ell^{2}}+\frac{a^2r_h^{2}}{\ell^{2}}\right)
\end{eqnarray}
The rest of the thermodynamic quantities can be written as
\begin{align}
S
&=
\frac{\pi \left(r_h^{2}+2 b r_h + a^{2}\right)}{G \Xi},
\\[6pt]
\Omega
&=
\frac{a \Xi}{r_h^{2}+2 b r_h + a^{2}}
+\frac{a}{\ell^{2}},
\\[6pt]
\Phi
&=
\frac{q r_h}{r_h^{2}+2 b r_h + a^{2}} .
\end{align}
Here $m$, $a$, and $b$ are, respectively, the mass, the spin parameter, and the dilaton charge. It is presumed that the presence of the dilaton charge parameter $b$ modifies the expressions for entropy ($S$), angular velocity ($\Omega$), and the electric potential ($\Phi$). The parameter $b$ is expressed as $b =q^2/2m$ where $q$ is the electric charge of the black hole. In the extended phase space of black hole mechanics, the cosmological constant is treated as the thermodynamic pressure through the relation $$P=-\frac{\Lambda}{8\pi}=\frac{3}{8\pi \ell^2}$$
The volume conjugate to the pressure is defined as.
$$V=\left(\frac{\partial M}{\partial P}\right)_{S,J,Q}.$$ With these definitions of the pressure and volume, we write the first law of the Kerr-Sen AdS black hole mechanics as
\[
dM = T\,dS + \Omega\,dJ + \Phi\,dQ+ PdV
\]
The corresponding Smarr relation could be obtained using the scaling arguments 
\[
M = 2T\,S + 2\Omega\,J + \Phi\,dQ-2P\,V
\]
The extended phase-space thermodynamics of the Kerr-Sen AdS black hole leads to van der Waals-like phase transition behaviour near the critical points. Such a notion could be used to define the topology of the thermodynamic defects for Kerr-Sen AdS black holes. \\
With the expressions of the thermodynamic quantities in hand, we are now ready to investigate the topological number of the four-dimensional Kerr-Sen AdS black hole. As discussed in Ref. \cite{
Wu:2020cgf,Gao:2021xtt}, one can first introduce the generalized off-shell free energy as
\begin{eqnarray}
\mathcal{F} = M -\frac{S}{\tau}, \,
\label{thermkerr}
\end{eqnarray}
for a black hole system with mass $M$ and entropy $S$, where $\tau$ is an extra variable that can be considered as the inverse temperature of the cavity surrounding the black hole. Only
When $\tau = 1/T$, the generalized free energy becomes on-shell.

As reported in Ref. \cite{Wu:2020cgf,Ali:2023jox}, a vector $\phi=(\phi^{r_h},\phi^\Theta)$ in two dimensions is conventionally written as
\begin{eqnarray}
\phi = \left(\frac{\partial \mathcal{F}}{\partial r_{h}}\, ,  -\cot\Theta\csc\Theta\right) \, ,
\label{phi}
\end{eqnarray}
where the parameter $0 \le \Theta \le +\infty$ is introduced for convenience and intuition. The
component $\phi^\Theta$ is divergent at $\Theta = 0, \pi$, and the direction of the vector
points outward there. Furthermore, as discussed in Ref. \cite{Wei:2022mzv}, the extremal points
of the generalized free energy landscape exactly corresponds to the on-shell black holes, so
the zero point of the component $\phi^{r_h}$ exactly meets the black hole solution. The
component $\phi^\Theta = 0$ eventually yields $\Theta = \pi/2$.

According to Duan's $\phi$-mapping topological current theory, a topological current can be
defined as \cite{Duan:1979ucg,Duan:1998kw,Fu:2000pb}
\begin{eqnarray}
\label{jmu}
j^{\mu}=\frac{1}{2\pi}\epsilon^{\mu\nu\rho}\epsilon_{ab}\partial_{\nu}n^{a}\partial_{\rho}n^{b}\, , \quad
\mu,\nu,\rho = 0,1,2,
\end{eqnarray}
where $\partial_{\nu} = \partial/\partial x^{\nu}$ and $x^{\nu}=(\tau,~r_h,~\Theta)$. The unit vector $n$ reads as
$n = (n^r, n^\Theta)$, where $n^r = \phi^{r_h}/{||\phi||}$ and $n^\Theta = \phi^{\Theta}/{||\phi||}$.
Once the topological current (\ref{jmu}) is constructed, one can easily argue that it is conserved; hence, the continuity equation $\partial_{\mu}j^{\mu} = 0$ is supposed to be valid identically pertaining to certain symmetry arguments. It is further shown that the topological current
$j^\mu$ is a $\delta$-function of the field configuration \cite{Duan:1998kw,Fu:2000pb,
Wei:2020rbh}
\begin{equation}
j^{\mu}=\delta^{2}(\phi)J^{\mu}\left(\frac{\phi}{x}\right)\, ,
\end{equation}
where the quantity $J^{\mu}\left(\phi/x\right)$ represents the three-dimensional Jacobian which has the form: $\epsilon^{ab}
J^{\mu}\left(\phi/x\right) = \epsilon^{\mu\nu\rho}\partial_{\nu}\phi^a\partial_{\rho}\phi^b$. It can be easily shown that the topological current $j^\mu$ identically vanishes except at the points when $\phi^a(x_i)$ = 0. This leads us to express the topological number $W$, in a certain parametric region $\Sigma$ as:
\begin{equation}
W = \int_{\Sigma}j^{0}d^2x = \sum_{i=1}^{N}\beta_{i}\eta_{i} = \sum_{i=1}^{N}w_{i}\, ,
\end{equation}
where $\beta_i$ is the positive Hopf index counting the number of loops of the vector field $\phi^a$
in the $\phi$-space whereas $x^{\mu}$ are evaluated around the zero points $z_i$. The parameter $\eta_{i}=\text{sign}(J^{0}
({\phi}/{x})_{z_i})=\pm 1$ represents the Brouwer degree and $w_{i}$ is the winding number for the $i$th
zero point of $\phi$ and is contained in the parameter region $\Sigma$. It is worth noting that if two loops $\Sigma_1$
and $\Sigma_2$ surround the same zero point of $\phi$, then they have the same winding number. On the other hand, if there is no zero point in the enclosed region, then one can have $W = 0$. The values of the $w_i$ play the crucial role in determining the local thermodynamic stability of a black hole. In this respect, one can also speculate that the positive or negative heat capacity of a black hole state corresponds to, respectively, the winding number $w=+1$ or $w=-1$, thereby reflecting the local stability or instability of the black hole state. As a specific rule to determine the difference between the counts of the thermally stable or unstable phases at a specific off-shell configuration, we need to evaluate the global topological number $W$. Hence, the number $W$ is a characteristic feature for each black hole system and is invoked as a classification parameter.\\
\subsection{Topological classes of Kerr-Sen AdS black hole}
In this section, we study the Kerr-Sen AdS black hole topology \cite{Wu:2020cgf}. From the results already given in Eq. (\ref{thermkerr}), one can easily obtain the
generalized free energy

\begin{equation}
\begin{aligned}
\mathcal{F}
&=
\frac{
3\left(a^{2} + 2 b r + r^{2}\right)
\left[
3 \tau
+ \pi r \left(
16 a^{2} P \pi
- 6
+ 16 b P \tau
+ 8 P r \tau
\right)
\right]
}{
2 G r \tau \left(3 - 8 a^{2} P \pi\right)^{2}
}
\end{aligned}
\label{offFree_E}
\end{equation}

\begin{figure}[t!]
\centering
\includegraphics[width=0.35\textwidth]{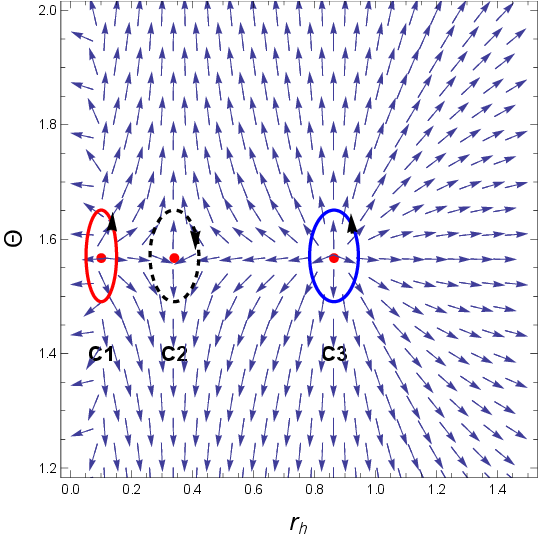}

\caption{Plot shows the unit vector field $n$
on the $r_h$--$\Theta$ plane for a four-dimensional Kerr-Sen AdS black hole with $\tau = 3.3$ and $a = 0.08$. The zero points (ZPs), marked with red dots, are located at
$(r_h,\Theta)=(0.1012204,\pi/2)$ , $(0.34086,\pi/2)$, and $(0.86402,\pi/2)$ . The contours $C_1$, $C_2$, and $C_3$ surrounding these zero points carry winding numbers $w_1 = 1$, $w_2 = -1$, and $w_3 = 1$, respectively, whose sum defines the total topological charge $W = 1$.}
\label{vector1}
\end{figure}
\begin{figure}[t]
\centering
\includegraphics[width=0.35\textwidth]{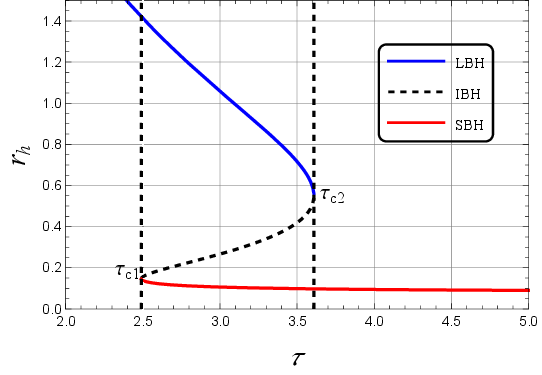}
\caption{The behavior of the zero points of $\phi^{r_h}$ in the $r_h-\tau$ plane for the Kerr-Sen AdS black hole. The blue solid, black dashed, and red solid curves correspond to the large black hole (LBH), intermediate black hole (IBH), and small black hole (SBH) branches, respectively, for the Kerr--Sen AdS black hole with $a = 0.08$, $b = 0.01$, and $P = 0.1193$. Two critical temperatures, $\tau_{c1} = 2.4903$ and $\tau_{c2} = 3.60951$, characterize the phase transitions.
}
\label{SchKerr}
\end{figure}

\begin{figure}[t!]
\centering
\includegraphics[width=0.35\textwidth]{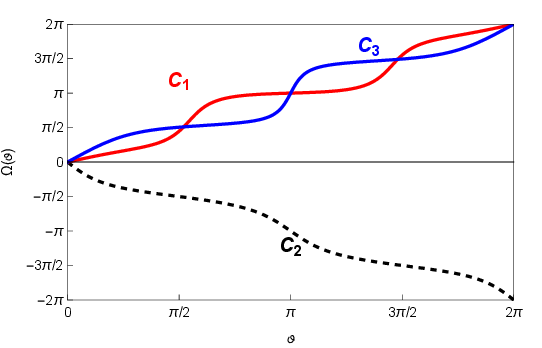}
\caption{The plot showing the behavior of $\Omega$ versus $\vartheta$ for the contours $C_1$ (Red solid curve),
$C_2$ (Black dashed curve), and $C_3$ (Blue solid curve) for the Ker--Sen AdS black hole.
\label{Kerrsen}}
\end{figure}

\begin{figure}[t!]
\centering
\includegraphics[width=0.35\textwidth]{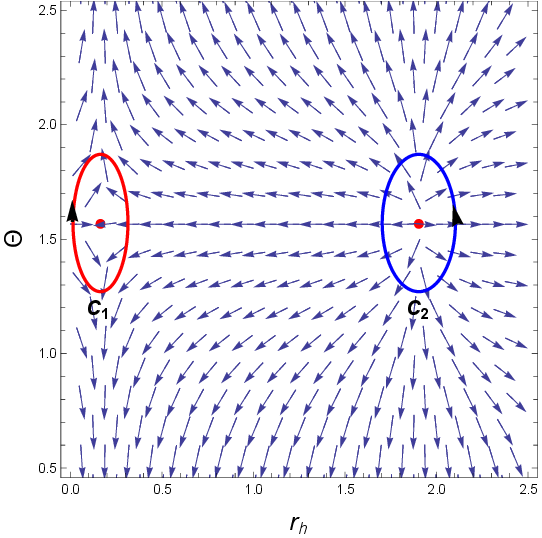}

\caption{The plot showing the unit vector field $n$
on the $r_h$-$\Theta$ plane for a four-dimensional Gibbons-Maeda-Garfinkle-Horowitz-Strominger (GMGHS) charged black holes in AdS spacetime with $\tau = 2$ when we switch off the rotation ($a = 0$). The zero points (ZPs), marked by red dots, are located at \((r_h,\Theta) = (0.16417,\pi/2)\) and \((1.903549,\pi/2)\). The orientations of the contours \(C_1\) and \(C_2\) are clockwise and anticlockwise, corresponding to the topological charges \(Q_{c_1} = -1\) and \(Q_{c_2} = +1\), respectively. The sum of the total topological charge $Q_\text{total} = 0$.}
\label{vector2}
\end{figure}

\begin{figure}[t!]
\centering
\includegraphics[width=0.35\textwidth]{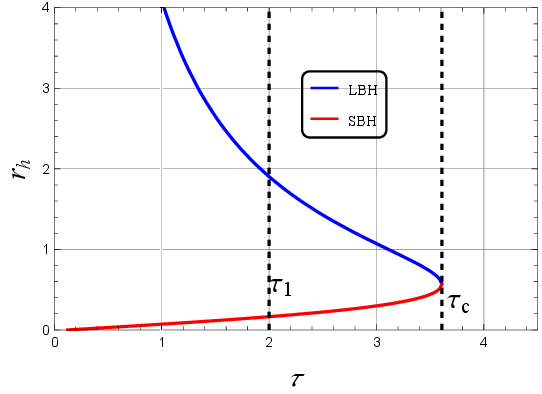}
\caption{The plot showing the behavior of the stable and unstable branches of the GMGHS AdS black hole phases in blue and red curves, respectively, for $a = 0,$ $b = 0.01$, and $P = 0.1193$. \label{4dKerr_01}}
\end{figure}

\begin{figure}[t!]
\centering
\includegraphics[width=0.35\textwidth]{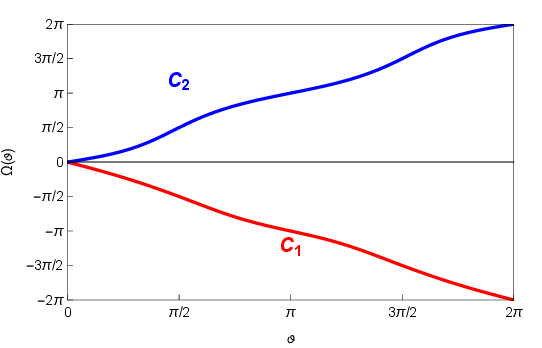}
\caption{The plot showing the behavior of $\Omega$ versus $\vartheta$ for the contours $C_1$ (red solid curve), and
$C_2$ (blue solid curve), for $a = 0,$ $b = 0.01$, and $P = 0.1193$. The blue and red curves correspond to the large black hole (LBH) and small black hole (SBH), respectively.
\label{windingspinless}}
\end{figure}
of the  Kerr--Sen AdS black hole. Then the components of the vector $\phi$ can be computed as

\begin{equation}
\phi^{r_h}
=
\frac{3}{2\,G_N\,(3 - 8 a^{2} P \pi)^{2}\,\tau}
\Bigg\{
\pi\bigl(32 a^{2} P \pi r - 12 r - 12 b\bigr)
+ \tau\Big[
\pi\bigl(32 b^{2} P + 8 a^{2} P + 64 b P r + 24 P r^{2}\bigr)
+ \frac{3(r^{2} - a^{2})}{r^{2}}
\Big]
\Bigg\}
\label{Phi_rh}
\end{equation}

\begin{equation}
\phi^{\Theta} = -\cot\Theta\,\csc\Theta
\end{equation}
By solving the equation $\phi^{r_h} = 0$, one can obtain its solution that is depicted by
a curve on the $r_h-\tau$ plane. For the four-dimensional Kerr-Sen AdS black hole, one can get
\begin{equation}
\tau
=
\frac{4 \pi (3-8 a^{2} P \pi) r^{3}}
{a^{2}(8 P \pi r^{2}-3)+r^{2}(3+32 b^{2} P \pi+64 b P \pi r+24 P \pi r^{2})}
\label{tau_ksads}
\end{equation}
The expressions given in Eqs.~(\ref{offFree_E}), (\ref{Phi_rh}), and (\ref{tau_ksads}) reduce to the corresponding expressions of the four-dimensional Kerr-AdS case when the dilation charge vanishes. From a topological viewpoint, it is well known that if a closed contour encloses a critical point, it gives a nonzero topological charge; otherwise, the topological charge vanishes. To calculate the topological charge, we construct three closed contours $C_1$, $C_2$, and $C_3$, parametrized by $\vartheta \in (0,2\pi)$ as
\begin{equation}
\begin{cases}
r = X \cos \vartheta + r_0, \\
\theta = Y \sin \vartheta + \dfrac{\pi}{2}.
\end{cases}
\label{contour}
\end{equation}
We choose $(X,Y,r_0)=(0.05,0.08,0.10122)$ for contour $C_1$, $(X,Y,r_0)=(0.08,0.08,0.3408)$ for contour $C_2$, and $(X,Y,r_0)=(0.08,0.08,0.864029)$ for contour $C_3$.
To better understand how the vector field behaves along each contour, we introduce a new quantity that quantifies the deflection of the vector field as--
\begin{equation}
\Omega(\vartheta)=\int_{0}^{\vartheta} \epsilon_{ab}\, n^{a}\,\partial_{\vartheta} n^{b}\, d\vartheta ,
\label{omega}
\end{equation}
where $n^a$ denotes the unit vector field and $\epsilon_{ab}$ is the two-dimensional Levi-Civita symbol. The topological charge is then given by
\begin{equation}
Q=\frac{1}{2\pi}\,\Omega(2\pi).
\label{charge}
\end{equation}

Fig.~\ref{vector1} shows the zero points of the vector field on a region of the $r_h$--$\Theta$ plane for the Kerr-Sen AdS black hole with parameters $a = 0.08$ and $b = 0.01$ at a constant pressure $P_0$. The blue arrows represent the vector field. The zero points, $\mathrm{ZP}_1$, $\mathrm{ZP}_2$, and $\mathrm{ZP}_3$, marked by red dots, are located at $(r_h,\Theta) = (0.10122,\,\pi/2)$, $(0.3408,\,\pi/2)$, and $(0.864029,\,\pi/2)$, respectively. The contours $C_1$, $C_2$, and $C_3$ are constructed to enclose the zero points. As a consequence, the winding numbers for the contours $C_1$ (red solid), $C_2$ (black dashed) and $C_3$ (blue solid), are identified as $w_1=+1$, $w_2=-1$, and $w_3=+1$, respectively. Regarding the global topological number, one can easily get $W=w_1+w_2+w_3=1$. This feature exactly mimics the RN-AdS and Kerr-AdS cases in four dimensions, where the global topological number is $+1$ \cite{Wei:2022dzw, Wu:2023sue}. This way, we can infer that for certain parametric values, the Kerr-Sen AdS black hole has the same geometric topology, and they belong to the same topological classes as far as the thermodynamic topology is concerned. However, the topological number of the Schwarzschild-AdS black hole is $0$; hence, we can infer that rotation and dilaton charge parameters play an important role in determining the topological number for Kerr-Sen AdS black holes.\\
The zero points $\phi^{r_h}$ in the $r_h-\tau$ plane are depicted in Fig.~\ref{SchKerr} for a fixed set of values of the rotation parameter, the dilaton charge, and the pressure, respectively, as  $a = 0.08$, $b = 0.01$, and $P = 0.1193$. With these values of the $a$, $b$ and $P$, we have two inverse critical temperatures, $\tau_{c1} = 2.4903$ and $\tau_{c2} = 3.60951$, characterize the phase transition behaviors. The parameter $\tau_{c1}$ corresponds to the creation point, whereas $\tau_{c2}$ accounts for the annihilation point. As shown in the Fig.~\ref{SchKerr}, we can have one large black hole branch for $\tau<\tau_{c1}$, three branches for $\tau_{c1}<\tau<\tau_{c2}$ and one small black hole branch for $\tau>\tau_{c2}$.\\
For the contours $C_1$, $C_2$ and $C_3$ , we plot $\Omega(\vartheta)$ in Fig.~\ref{Kerrsen}. The contours $C_1$ and $C_3$ evolve in the positive (anti-clockwise) $\Omega$ direction, while the contour $C_2$ evolves in the negative (clockwise) $\Omega$ direction. This is due to the difference in orientations, where winding numbers (or topological charges) of opposite signs are produced. This indicates different topological properties for various thermodynamic phases of black holes. The fact that the contours remain well separated and do not overlap shows that the zero points are topologically protected. As a result, one type of phase transition behavior cannot smoothly evolve into another.  Any transition between them requires the creation or annihilation of the zero point, indicating a topological phase transition of the black hole.

As a result, for the contour $C_1$ and $C_3$, $\Omega(\vartheta)$ increases monotonically and approaches $2\pi$ at $\vartheta=2\pi$. Therefore, the topological charge associated with the critical point $\mathrm{CP}_1$ and $\mathrm{CP}_3$ is
\begin{equation}
Q_{\mathrm{CP}_1} = Q_{\mathrm{CP}_3}=+1
\end{equation}
and hence corresponds to the generation of topological charges on the contours. Further, for the contour $C_2$, $\Omega(\vartheta)$ decreases monotonically and approaches $-2\pi$ at $\vartheta=2\pi$. Therefore, the topological charge associated with the critical point $\mathrm{CP}_2$ is
\begin{equation}
Q_{\mathrm{CP}_2}=-1.
\end{equation} 
and we have one annihilation point there. According to our classification, the presence of three critical points implies that the system admits three distinct thermodynamic phases, namely small, intermediate, and large black holes. Each critical point marks a first-order phase transition 
between two neighboring phases. Consequently, the system undergoes small--intermediate and intermediate--large black hole phase transitions in different regions of the parameter space, resulting in a rich and complex phase structure. The corresponding charges are denoted as $Q_{\mathrm{CP}_1}$ , $Q_{\mathrm{CP}_2}$ and  $Q_{\mathrm{CP}_3}$, wherein the total charge is encoded to be
\begin{equation}
Q_{\mathrm{tot}} = Q_{\mathrm{CP}_1} + Q_{\mathrm{CP}_2} + Q_{\mathrm{CP}_3} = 1
\end{equation}
The plot in Fig.~\ref{vector2} shows the behavior of the unit vector fields $n$ in the parametric space $(r_h$,$\Theta$) for the four-dimensional GMGHS charged black holes in AdS spacetime at inverse temperature $\tau = 2$ in the limit of vanishing rotation parameter ($a = 0$). The ZPs enclosed by the contours are marked by red dots and are located at \((r_h,\Theta) = (0.16417,\pi/2)\) for $C_1$ and \((1.903549,\pi/2)\) for $C_2$. Depending on the orientations of the contours \(C_1\) and \(C_2\), we define the topological charges as \(Q_{c_1} = -1\) and \(Q_{c_2} = +1\), respectively, so that the sum of the total topological charge is $Q_\text{total} = 0$. To see how the black hole evolves, we plot in Fig.~\ref{4dKerr_01}, the zero points of $\phi^{r_h}$ in the parameter space ($r_h,\tau$) for a set of values of $a = 0,$ $b = 0.01$, and $P = 0.1193$. In this case, we only get one inverse critical temperature, $\tau_{c} = 3.6$, characterizing the phase transition behavior of small to large black hole phases. Such types of phase transitions are alien to the charged AdS black holes, which exhibit three distinct phases: SBH, IBH, and LBH. \\

For two branches of the extended phase space thermodynamics, we get the small and large black hole phases with the winding numbers ($``w=+1"$) and ($``w=-1"
$), respectively, when the spin parameter is set to zero. The values of the topological charges obtained from the integrals are shown in Fig.~\ref{windingspinless}. Consequently, at the values of the parameters $a= 0,$ $b = 0.01$, and $P = 0.1193$, we have a vanishing total topological charge $Q_\text{total}=0$ resulting from the contributions of the zero points and thereby canceling each other. A topological transition is indicated by the emergence of such a pair, in which zero points are formed or destroyed while maintaining the conservation of the total topological charge.\\

\begin{figure}[t!]
\centering

\subfloat[]{%
\includegraphics[width=0.35\linewidth]{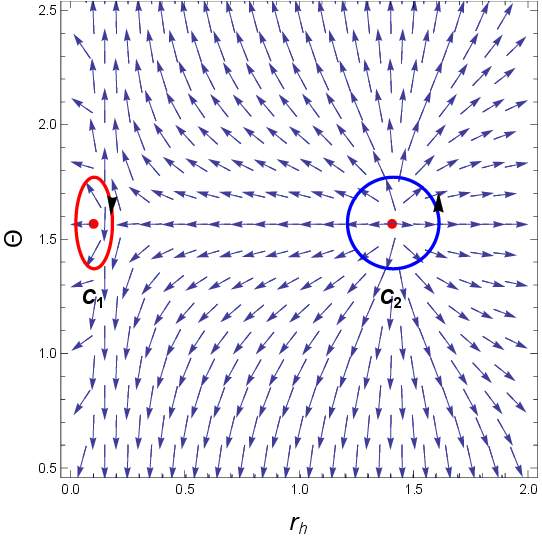}
}
\hspace{0.04\textwidth}
\subfloat[]{%
\includegraphics[width=0.35\linewidth]{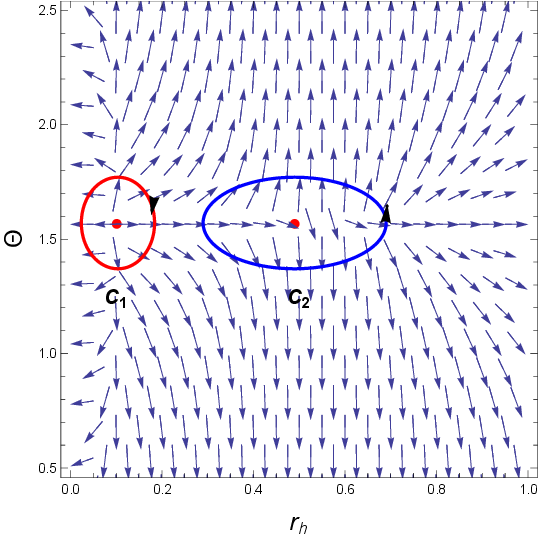}
}

\caption{
Unit vector field $n$ on the $r_h$--$\Theta$ plane for a four-dimensional Kerr--Sen AdS black hole
with rotation parameter $a=0.08$,Pressure $P= 0.1193$ and dilatonic charge $b= $0.01 :
(a) $\tau=\tau_{c1}$ The zero points (ZPs), marked by red dots, are located at
$(r_h,\Theta)=(1.4208,\pi/2)$.
The contours $C_1$ and $C_2$, enclosing the zero points, carry winding number
$w_1=0$ and $w_2=1$, respectively, and (b) $\tau=\tau_{c2}$.
The zero points (ZPs), marked by red dots, are located at
$(r_h,\Theta)=(0.10122,\pi/2)$.
The $C_1$ and $C_2$ are two closed loops, $C_1$  enclose the zero point, while the $C_2$ does not. The contours carry winding numbers
$w_1=1$ and $w_2=0$, respectively.
Their sum gives the total topological charge $W=1$.
}
\label{vectoratcritical}
\end{figure}

\begin{figure}[t]
\centering

\subfloat[]{%
\includegraphics[width=0.35\linewidth]{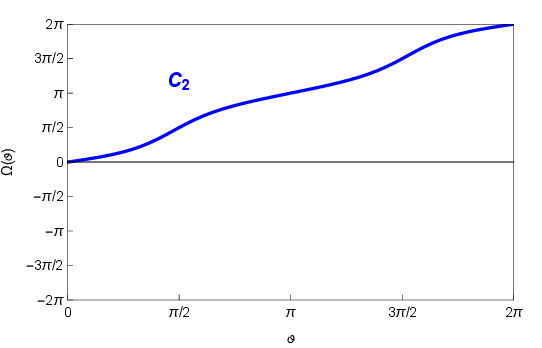}
}
\subfloat[]{%
\includegraphics[width=0.35\linewidth]{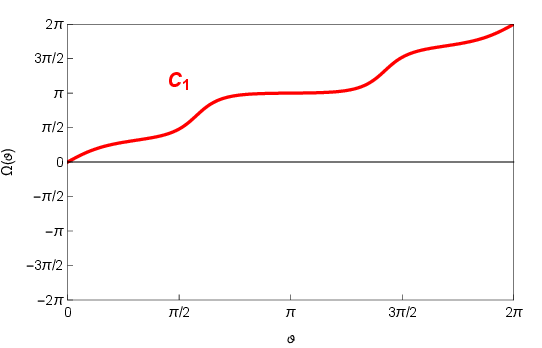}
}

\caption{
Plots of $\Omega$ versus $\vartheta$ for the contours
$C_1$ (red solid curve) and $C_2$ (blue solid curve)
for the Kerr--Sen AdS black hole.
Panel (a) corresponds to Fig.~\ref{vectoratcritical} (a),
while panel (b) corresponds to Fig.~\ref{vectoratcritical} (b).
}

\label{fig:OmegaVartheta}
\end{figure}

Figs.~\ref{vectoratcritical} and \ref{fig:OmegaVartheta} show the topological nature of the thermodynamic phase structure of the Kerr--Sen AdS black hole. At the critical values $\tau=\tau_{c1}$ and $\tau=\tau_{c2}$, isolated zero points (ZPs) emerge in the vector field, signaling thermodynamic criticality. Although the location of the zero point shifts from the large-$r_h$ regime to the small-$r_h$ regime as $\tau$ varies, the total topological number remains invariant and equals to $W=1$. This invariance demonstrates that the phase transition is topological in origin and that the critical behavior is protected against smooth deformations of the thermodynamic parameters. \\
\begin{table}[h]
\centering
\begin{tabular}{l|c|c|c|c|c}
\hline\hline
        & $I_1$ & $I_2$ & $I_3$ & $I_4$ & $W$ \\
\hline
Case I   & $\leftarrow$ & $\uparrow$ & $\rightarrow$ & $\downarrow$ & $-1$ \\
Case II  & $\leftarrow$ & $\uparrow$ & $\leftarrow$  & $\downarrow$ & $0$  \\
Case III & $\rightarrow$ & $\uparrow$ & $\leftarrow$  & $\downarrow$ & $+1$ \\
Case IV  & $\rightarrow$ & $\uparrow$ & $\rightarrow$ & $\downarrow$ & $0$  \\
\hline\hline
\end{tabular}
\caption{The directions of the arrows along the four segments $I_1$--$I_4$ and the corresponding topological number $W$ for each configuration.}
\end{table}

\begin{figure}[t!]
\centering
\includegraphics[width=0.35\textwidth]{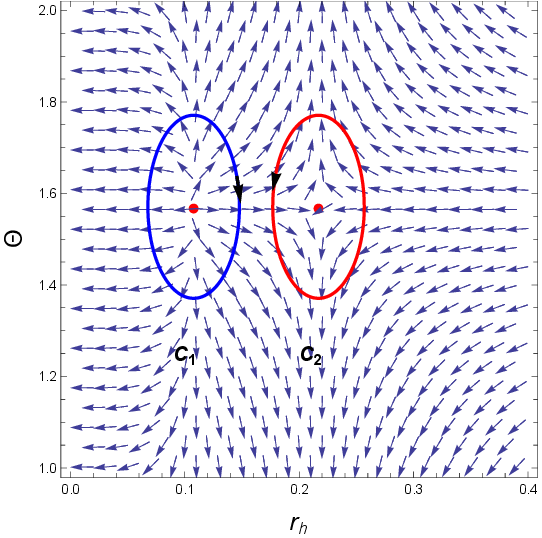}

\caption{Plot shows the unit vector field $n$
on the $r_h$--$\Theta$ plane for a four-dimensional Kerr-Sen black hole with
$\tau = 3.3$ and $a = 0.08001$.
The zero points (ZPs), marked by red dots, are located at
\((r_h,\Theta) = (0.107715,\pi/2)\) and \((0.216858,\pi/2)\).
The orientations of the contours \(C_1\) and \(C_2\) are clockwise and
anticlockwise, corresponding to the topological charges
\(Q_{c_1} = +1\) and \(Q_{c_2} = -1\), respectively.
The sum of  total
topological charge $Q_{total} = 0$.}
\label{vector3}
\end{figure}

\begin{figure}[t!]
\centering
\includegraphics[width=0.35\textwidth]{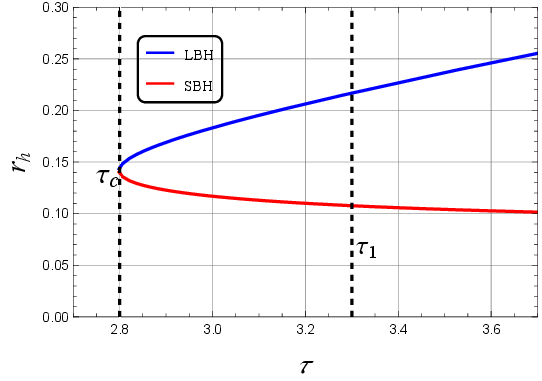} 
\caption{The plot shows the stable and unstable black hole branches in blue and red, respectively, for the Kerr--Sen black hole at $a= 0.08$ and $b=0.01$. The blue and red curves correspond to the large black hole (LBH) and small black hole (SBH), respectively\label{taurkerrsenn}}.
\end{figure} 

\begin{figure}[t!]
\centering
\includegraphics[width=0.35\textwidth]{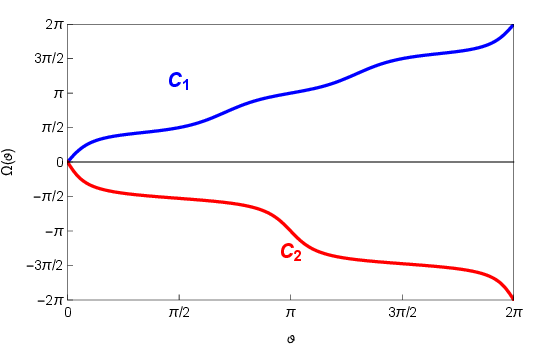}
\caption{The plot of $\Omega$ versus $\vartheta$ for the contours $C_2$ (Red s curve),
 and $C_1$ (Blue Solid curve) for the $\Lambda  = 0$.
\label{windingkerrsen}}
\end{figure}

Next, we consider the case when the negative cosmological constant vanishes, i.e., when $\Lambda=0$. The plot in Fig.~\ref{vector3} shows the distribution of the unit vector field $n$ on the $r_h–\Theta$ plane of a four-dimensional Kerr-Sen black hole with the ensemble temperature $\tau=3.3$ in the off-shell configuration. The rotation and dilaton charge parameters are set to $a = 0.08001$ and $b = 0.01$. The ZPs, marked by red dots, are enclosed by the contours $C_1$ and $C_2$ and are found to be located at $(r_h, \Theta)= (0.107715,\pi/2)$ and $(0.216858,\pi/2)$. Considering the orientation of the contours, $C_1$ and $C_2$, which are in the clockwise and anticlockwise directions. The contours $C_1$ and $C_2$, correspond to topological charges \(Q_{C_1} = +1\) and \(Q_{C_2} = -1\), respectively. This amounts to the total topological charge $Q_\text{total} = 0$. On the other hand, in Fig.~\ref{taurkerrsenn} we demonstrate the zero points of the component $\phi^{r_h}$ in the $(r_h-\tau)$ plane for the Kerr-Sen black hole with a fixed set of values of the rotation and dilaton charge parameters as $a=0.08$ and $b=0.01$. At the value of $\tau=\tau_1$, we have two intersection points of the Kerr-Sen black hole. This behavior is similar to the cases of the RN and Kerr black hole configurations. At the intersection points, one has the condition $\tau=1/T$, and we have on-shell black hole solutions. Obviously, the intersection points coincide with each other at the point when $\tau=\tau_c$, and when $\tau<\tau_c$ they vanish. One can easily check that at $\tau=\tau_c$, we have $$\frac{d^2\tau}{dr_h^2}>0$$
indicating a creation point. The creation point at $\tau=\tau_c$ is accountable for the partitions of the Kerr-Sen black holes into upper and lower branches with their corresponding winding numbers $w=-1$ and $w=+1$, respectively. \\
Further, in Fig.~\ref{windingkerrsen}, we illustrate the behavior of the topological number calculation by plotting the $\Omega(\vartheta)$ for the contours in Fig.~\ref{vector3}. As the contours $C_1$ and $C_2$ do not intersect, the ZPs are topologically protected. The opposite orientations of these contours lead us to have the winding numbers bearing the opposite signs. While the red solid curve in Fig.~\ref{windingkerrsen} carries the negative topological charge, the blue solid curve represents the positive topological charge. Thus, the total topological charge becomes zero.    
\section{Kerr--Sen topology using complex residue techniques}
\label{III}
In this section, we evaluate the topological number using the complex residue technique \cite{Fang:2022rsb}. In this method, we need to determine the winding number ($w_i$) for each pole arising from the residue method. Adding these winding numbers corresponding to the poles, we can get the total topological charge. Inspired by this, we first solve $\partial_{r_h}\mathcal{F}=0$
to get the expression for $\tau$ as a function of the horizon radius
\begin{equation}
\tau=\mathcal{G}(r_h),
\end{equation}, 
The next step would be to replace the horizon radius in terms of the complex variable $z$ such that \cite{Fang:2022rsb,Hazarika:2024dex}$$\tau(r_h\to z)=\mathcal{G}(z)$$
The substitution of the event horizon radius ($r_h$) with a complex variable $z$ is meant for the fact that the real variable is now translated to the complex plane $\mathbb{C}$ with a complex number given as $z=x+ i y$ where $x, y\in \mathbb{R}$ are real variables. We aim to generate a two-dimensional parametric space where we can define a real contour. This contour must enclose the thermodynamic defects that are determined from the expressions of $\mathcal{G}(z)$.\\
Now we define a characterized complex function $\mathcal{R}(z)$ as follows
\begin{equation}\label{cf}
\mathcal{R}(z)\equiv\frac{1}{\tau-\mathcal{G}(z)},
\end{equation}
The characterized complex function $\mathcal{R}(z)$ is related to the generalized free energy. To be more specific, the complex residue method is a byproduct of the contour integral of a rational complex function $\mathcal{R}(z)$ that, except at the isolated singular points $z_1, z_2, \ldots, z_m$ interior to $C$, is analytic. The contour $C$ is a simple closed curve oriented in an anti-clockwise direction. As usual, the complex analytic function is defined within and on the contour $C$. The application of the residue theorem follows this, and then the computation of the integral of the function $R(z)$ over the path $C_k$ surrounding $z_k$, which is accordingly written as,

\begin{equation}
\oint_C \frac{\mathcal{R}(z)}{2\pi i} d z= \sum_{k=1}^m \operatorname{Res}\left[\mathcal{R}\left(z_k\right)\right].
\end{equation}
In this way, the previously defined topological defects by $\phi^{r_h}=0$ ( as shown in Eq. \ref{phi}) on the real plane are now transformed into isolated singular points of the rational complex function $\mathcal{R}(z)$ on the $z$-plane. \\
Since each singular point identified as a thermodynamic defect can now be enclosed with an arbitrary contour. As the characterized rational function $\mathcal{R}(z)$ is analytic everywhere except at the singular points $z_1,z_2,...,z_N$. Our task would be to find out the integral of the function $\mathcal{R}(z)$ using the residue theorem along each closed contour enclosing the corresponding singular point. As stated by the fundamental postulate of the Cauchy-Goursat theorem, the integral of the complex function $\mathcal{R}(z)/2\pi i$ encircling all the singular points by the contour $C$ is equivalently written as the integrals of $\mathcal{R}(z)$ along $C_i$ in the interior enclosing $z_i$.\\
Subsequently, we can characterize the winding numbers by calculating the residue of the complex function $\mathcal{R}(z)$ at its pole points, which are similar to the topological defects related to the on-shell black hole configurations. In what follows, to determine the topological characteristics of these defects, we must normalize the residue, thereby defining the winding number $w_i$ associated with a singular point $z=z_i$, so that \cite{Fang:2022rsb}
\begin{equation}\label{wind}
w_i =\frac{\textrm{Res}\mathcal{R}(z_i)}{|\textrm{Res}\mathcal{R}(z_i)|}=\textrm{sgn}[\textrm{Res}\mathcal{R}(z_i)],
\end{equation}
where $|\textrm{Res}\mathcal{R}(z_i)|$ is the absolute value of the residue of the complex-valued function, i.e., $\textrm{Res}\mathcal{R}(z_i)$. The quantity $\textrm{sgn}(x)$ is known as the sign function, which has the following functional properties
\[
\mathrm{sgn}(x)=
\begin{cases}
+1, & x>0, \\
0, & x=0, \\
-1, & x<0.
\end{cases}
\]

It is certainly worthy of mentioning that the singular points of our interest are all real, hence satisfying the condition $\textrm{Res}\mathcal{R}(z_i)\in\mathbb{R}$. Therefore, the global topological number $W$ of the black hole spacetime is determined to be
\begin{equation}\label{topw}
W=\sum_i w_i.
\end{equation}
This can be understood by applying the Cauchy-Goursat theorem, which states that the integral of a complex function over an exterior contour is evaluated through the integrals along simple closed contours enclosing the isolated singular points. Then, we have
\begin{equation}
\label{caug}
\oint_C \mathcal{R}(z) d z= \sum_i \oint_{C_i} \mathcal{R}(z) d z.
\end{equation}

\begin{figure}[httb!]
\centering
\includegraphics[width=0.5\textwidth]{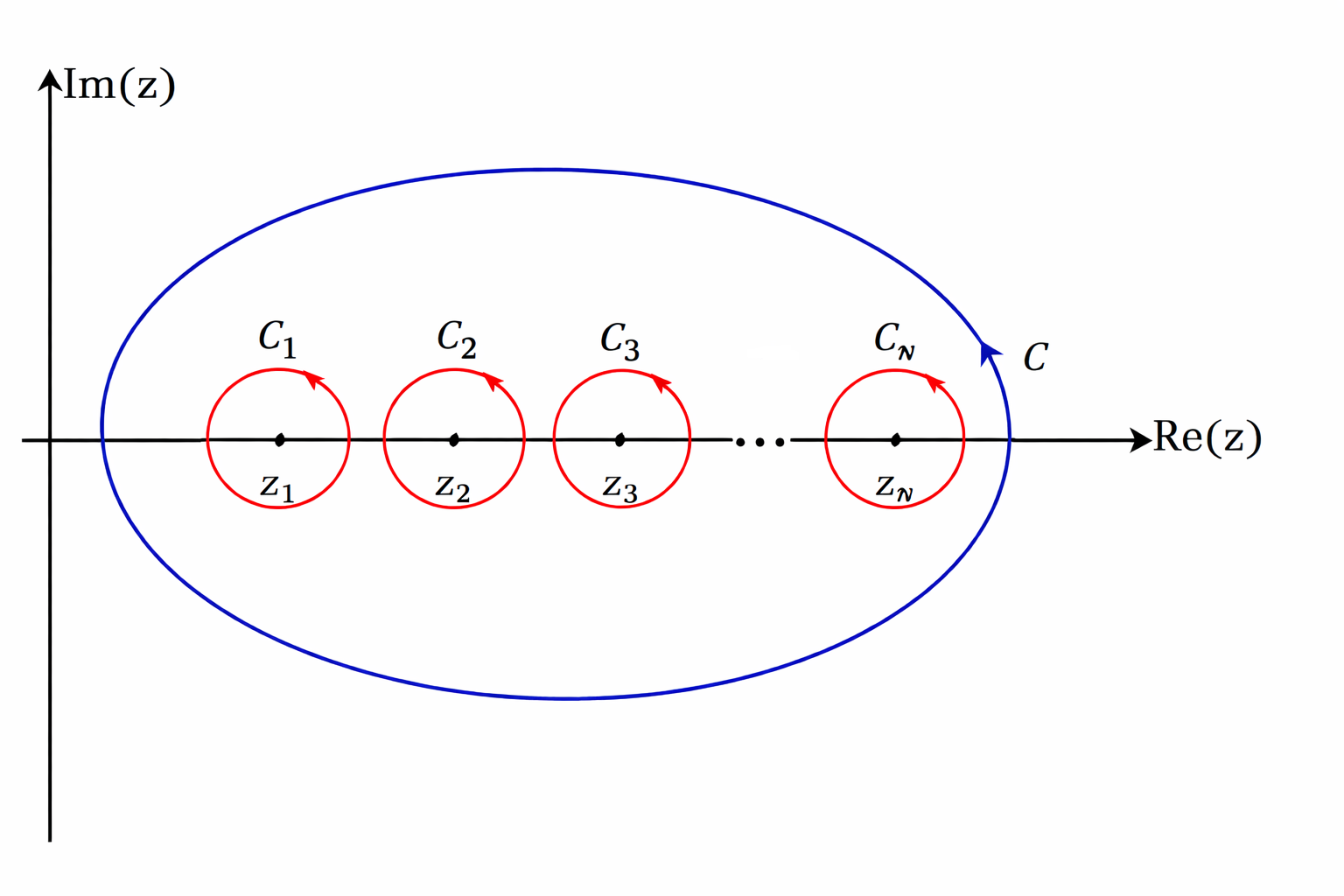}
\caption{A schematic diagram illustrating the contour integral of a function that is analytic everywhere except at a finite number of isolated singular points $z_1, z_2, z_3, \ldots, z_N$, enclosed by a global contour $C$ and a set of individual contours $C_1, C_2, C_3, \ldots, C_N$ surrounding each singularity.}
\label{vector2}
\end{figure}

\begin{figure}[t]
\centering

\subfloat[]{%
\includegraphics[width=0.33\linewidth]{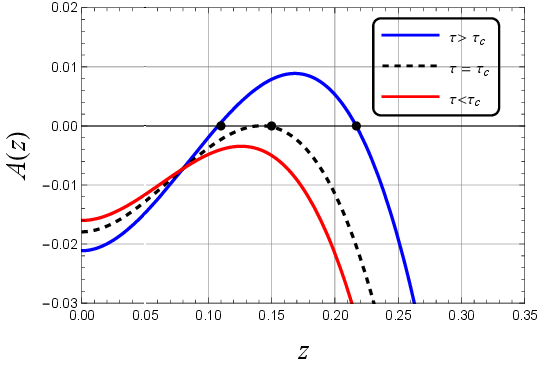}}
\hfill
\subfloat[]{%
\includegraphics[width=0.33\linewidth]{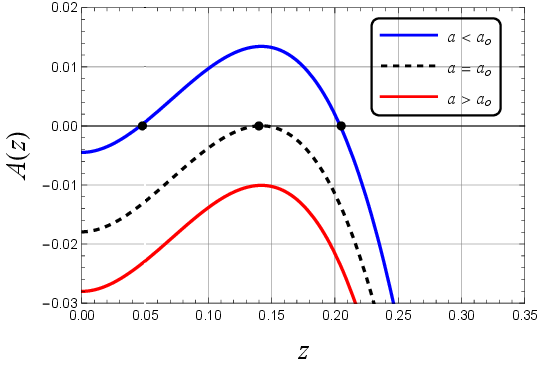}}
\hfill
\subfloat[]{%
\includegraphics[width=0.33\linewidth]{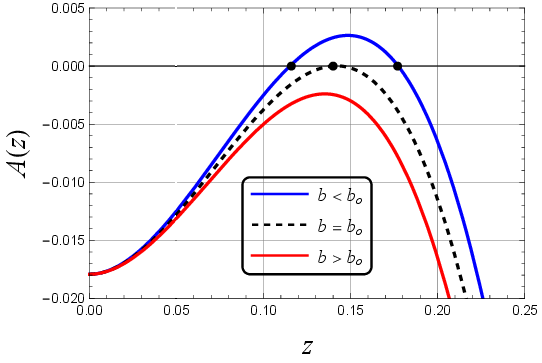}}

\caption{Roots of the polynomial equation $A(z)=0$ for different values of the
model parameters. Panels (a), (b), and (c) correspond to variations in the
parameters $\tau$, $a$, and $b$, respectively. The evolution of the roots
illustrates the dependence of the solution structure on the underlying
physical parameters.}
\label{fig:roots_polynomial}
\end{figure}
The complex residue technique provides us with an effective framework to satisfactorily classify and analyze various isolated singular points. Of particular interest, it can enable us to calculate the winding numbers for higher-order poles, which illustrate the features like creation, annihilation, and the critical points. In this work, we demonstrate such possibilities to compute the thermodynamic topologies of the Kerr-Sen black hole solutions.\\

When $\Lambda$ $=0$, the Kerr-Sen AdS black hole becomes Kerr-Sen (KS) such that,
\begin{equation}
\Delta_{r}
=r^{2}+2br+a^{2}-2mr,
\end{equation}
The generalized off-shell free energy $\mathcal{F}$, of the Kerr-Sen black hole system from Eq.(\ref{thermkerr}), is expressed as
\begin{equation}
\mathcal{F} = \frac{(a^{2} + 2 b r + r^{2})\,(\tau-2\pi r )}{2 r \tau}
\end{equation}
As defined in Eq.~\ref{cf}, the characterized rational complex function of the Kerr-Sen black hole is computed to be
\begin{equation}
\mathcal{R}_\text{KS} =
\frac{a^{2}-z^{2}}{4\pi z^{2}(b+z)-(z^{2}-a^{2})\,\tau}
\end{equation}
Since the denominator is a cubic function of $z$, we can rewrite the above equation in the form
\begin{equation}
\mathcal{R}_\text{KS}(z) =
\frac{a^{2}-z^{2}}{4\pi (z-z_1)(z-z_2)(z-z_3)}= \frac{a^{2}-z^{2}}{\mathcal{A}(z)}
\end{equation}
where the polynomial function $\mathcal{A}(z)$ is expressed as
\begin{equation}
\mathcal{A}(z) =4\pi z^2 (b+z)-(z^2-a^2)\tau
\end{equation}
We need to solve this polynomial by equating it to zero for different values of the parameters $a$, $b$, and $\tau$. Once we solve for $z$, we can distinguish the winding numbers using Eq.~(\ref{wind}). \\
We demonstrate in Fig.~\ref{fig:roots_polynomial} the plot of the complex polynomial $\mathcal{A}(z)$ for three different cases. Fig.~\ref{fig:roots_polynomial}a shows the behavior of $\mathcal{A}(z)$ for different values of the parameter $\tau$ with $a= 0.08 $ and $b= 0.01$. There is only one pole for $\tau=\tau_c$, two poles for $\tau>\tau_c$, and no pole for $\tau<\tau_c$. For $\tau=\tau_c$, we have the large black hole branch, and the pole is located at $z= 0.141802$ as shown by the black dashed curve. The winding number corresponding to the pole at $z= 0.141802$ is assigned as $w=+1$. While for $\tau>\tau_c$, we have two poles located at $z_1= 0.107716$ and $z_2= 0.216858$, representing the small and large black hole phases. The corresponding winding numbers around the poles at $z_1$ and $z_2$ are represented as $w_1=-1$ and $w_2=+1$. Hence, the global topological number is calculated to be $W=w_1+w_2=0$. For $\tau<\tau_c$, we do not have any pole and hence no topological charge. Similar situations are observed in Figs.~\ref{fig:roots_polynomial}b and~\ref{fig:roots_polynomial}c, where we take the variations of the rotation parameter $a$ and the dilaton charge parameter $b$ for a fixed value of $\tau$, respectively. In Fig.~\ref{fig:roots_polynomial}b, we show the single pole ($a=a_0$), the double poles ($a>a_0$), and no pole condition ($a<a_0$) for $\tau= 2.79857$. Similarly, in Fig.~\ref{fig:roots_polynomial}c, we depict the single, double, and no pole conditions for $b=b_0$, $b>b_0$, and $b<b_0$, respectively, at a fixed $\tau= 2.79857$. 

\section{Conclusion}
\label{IV}

The Topology has emerged as a powerful tool for understanding black hole thermodynamics. By treating black hole phases as topological defects in a thermodynamic space, we can categorise them in a way that doesn't depend on coordinates -- revealing universal features that might otherwise stay hidden. The Kerr-Sen black hole provides an incredibly interesting setting for exploring these ideas. It comes from string theory, it rotates, and it carries a dilaton charge—ingredients that could potentially change the topological picture we've seen in simpler black holes. Motivated by this, we set out to see how rotation and the dilaton field influence thermodynamic topology and whether the patterns established for Schwarzschild, Reissner-Nordström, and Kerr black holes still hold in this richer, more realistic setting.

Indeed, we studied the thermodynamics of Kerr-Sen AdS black holes using two different approaches to understand their topological properties. The first method uses Duan's $\phi$-mapping theory, while the second is a new technique based on complex residues. By looking at the off-shell free energy and the vector field it creates in the space of parameters $(r_h, \Theta)$, we examined three versions of the black hole: the full Kerr-Sen AdS black hole, its non-rotating version called GMGHS AdS (where $a = 0$), and the flat space version with no cosmological constant (where $\Lambda = 0$). We note that the Kerr-Sen AdS black hole exhibits three thermodynamic phases, viz., small, intermediate, and large black holes. These phases have winding numbers $+1$, $-1$, and $+1$ respectively. When we add these up, we get a total topological charge of $Q = +1$. It puts the Kerr-Sen AdS black hole in the same topological family as the well-known Reissner-Nordstr\"om AdS and Kerr-AdS black holes. What's really interesting is that changing the dilaton charge $b$ doesn't affect the total topological charge at all—it makes no difference. This tells us that the dilaton field doesn't really change the black hole's basic topological character. But the rotation parameter $a$ is a whole different story. It plays a major role in shaping the phase structure and determining how many critical points appear. So when it comes to thermodynamic topology, spin matters a lot more than the dilaton charge does.

Looking at the special cases gave us some interesting results. When we turn off rotation ($a = 0$) to get the GMGHS AdS black hole, the total topological charge becomes zero ($W = 0$). It happens because we have two phases with winding numbers $+1$ and $-1$ that cancel each other out. The same thing happens for the flat Kerr-Sen black hole (with $\Lambda = 0$)---it also gives $W = 0$ even though it has two distinct phases. What this tells us is that while rotation and the cosmological constant do affect the topology, the dilaton charge doesn't change the overall topological picture. It hints at some universality in how black hole thermodynamics behaves topologically. One of the most interesting parts of this work is the new complex residue method we used. Rather than Duan's approach, we took the thermodynamic functions and extended them into the complex plane. We built a complex function $\mathcal{R}(z) = 1/\tau - \mathcal{G}(z)$ and found that its singular points match exactly with the thermodynamic defects. The residues at these points give us the winding numbers through the simple formula $w_i = \mathrm{sgn}[\mathrm{Res}\mathcal{R}(z_i)]$. This is a much cleaner mathematical approach that yields the same results as Duan's method. It also gives us a fresh perspective on critical points and might help us understand phenomena such as creation and annihilation events in phase transitions.

When we looked at the complex polynomial $\mathcal{A}(z) = 4\pi z^2(b+z) - (z^2 - a^2)\tau$ for the flat space Kerr-Sen black hole, we saw three different situations. At the critical temperature $\tau = \tau_c$, there's just one pole. For temperatures above $\tau_c$, we get two poles corresponding to small and large black hole phases with winding numbers $-1$ and $+1$. And for temperatures below $\tau_c$, there are no poles at all. We saw similar patterns when we changed the rotation parameter $a$ and the dilaton charge $b$. This shows that the topological classification is quite robust and that the complex residue method works really well for capturing what's going on in black hole thermodynamics. Our results point to some bigger ideas. First, the fact that the dilaton charge doesn't change the total topological charge suggests that this kind of topological classification might be a universal feature of black hole thermodynamics -- it doesn't seem to care too much about certain details like matter couplings. Second, the fact that both methods give the same results is reassuring and helps us better understand what thermodynamic defects really mean mathematically. Third, using complex analysis in black hole thermodynamics opens new possibilities -- now we can apply powerful tools from complex analysis to study phase transitions.

Looking ahead, there's a lot more we could explore. For starters, we could try to see how the thermodynamic topology we've been studying connects with holographic dual field theories. That might help us understand how topological behaviour in gravity manifests as distinct phases in field theories. We could also take these ideas further and examine non-equilibrium situations—such as what happens when black holes form, change, or evaporate over time. Another interesting question is whether the topological patterns we've seen are universal—do they show up in other black hole solutions from string theory or higher-curvature gravity as well? And finally, digging into the topology might give us hints about what black holes look like at a microscopic level and what they can reveal about the deeper nature of quantum gravity.

To wrap up, this work shows that topological methods give us deep insights into black hole phase transitions. They provide a way of looking at things that doesn't depend on coordinates and reveals universal features of black hole thermodynamics. The complex residue method, in particular, is both powerful and elegant. It complements existing approaches and could be really useful for future studies on the role of topology in quantum gravity and holography.\\

\section*{Acknowledgements}
The authors would like to thank the Inter-University Centre for Astronomy and Astrophysics (IUCAA), Pune, for hospitality while part of this work was being done.
\bibliography{kersentopo}
\bibliographystyle{unsrt}

\end{document}